\documentclass[prl,showpacs,amsfonts,amssymb,floats,twocolumn,superscriptaddress,aps]{revtex4}

\usepackage[final,dvips]{epsfig}
\newcommand{\pd}{{\phantom{\dagger}}}

\begin{document}

\title[]
{Dissipative Exciton Transfer in Donor-Bridge-Acceptor Systems:
Numerical Renormalization Group Calculation of Equilibrium
Properties}

\author{Sabine Tornow}

\affiliation{\mbox{Theoretische Physik III, Elektronische
Korrelationen und Magnetismus, Universit\"at Augsburg, 86135
Augsburg, Germany}}
\author{Ning-Hua Tong \footnote{present adress: Department of Physics, Renmin University of China,
    Beijing 100872, China
}}
\affiliation{\mbox{Institut f\"ur Theorie der Kondensierten
Materie, Universit\"at Karlsruhe, 76128 Karlsruhe, Germany}}
\author{Ralf Bulla}
\affiliation{\mbox{Theoretische Physik III, Elektronische
Korrelationen und Magnetismus, Universit\"at Augsburg, 86135
Augsburg, Germany}}
%\date{Sep 2, 2003}
\date{\today}

\begin{abstract}
We present a detailed model study of exciton transfer processes in
donor-bridge-acceptor (DBA) systems. Using a model which includes
the intermolecular Coulomb interaction and the coupling to a
dissipative environment we calculate the phase diagram, the
absorption spectrum as well as dynamic equilibrium properties with
the numerical renormalization group. This method is
non-perturbative and therefore allows to cover the full parameter
space, especially the case when the intermolecular Coulomb
interaction is of the same order as the coupling to the
environment and perturbation theory cannot be applied. For DBA
systems up to six sites we found a transition to the localized
phase (self-trapping) depending on the coupling to the dissipative
environment. We discuss various criteria which favour delocalized
exciton transfer.
\end{abstract}

\pacs{
%05.10.Cc Renormalization Group Methods,
%34.70.+e Charge Transfer,
%85.65.+h Molecular Electronic Devices,
%82.39.Jn Charge (Electron, Proton) Transfer in Biological Systems,
%78.67.Hc Quantum Dots,
71.27.+a Strongly correlated electron systems,
71.35.Aa Frenkel Excitons and Self-Trapped Excitons,
71.35.Cc Intrinsic Properties of Excitons; Optical Absorption
Spectra.}

\maketitle
\section{Introduction}
Exciton transfer belongs to the key processes in many chemical and
biological systems, organic based nano\-structures and
semiconductors \cite{Hu,Agranovich,May}. The progress in
manufacturing molecular electronic devices, biological hybrid
systems, and model systems based on quantum dots, nanoscale
molecular aggregates and bio-engineered proteins opens the door to
understand these fundamental processes \cite{Fleming} and also to
find applications in (bio-) molecular electronics, biosensing, and
quantum computation \cite{Briggs}.

Excitons are electron-hole pairs which do not transfer charge but
energy by deexciting a donor molecule followed by the excitation
of an acceptor molecule. The radiationless excitation transfer is
caused by dipole and exchange interactions and proceeds via a
short lived virtual photon \cite{Schulten1}. In this work we
consider Frenkel excitons \cite{Davydov} where the exciton is a
molecular excitation with an electron in the lowest unoccupied
molecular orbital (LUMO) and a hole in the highest occupied
molecular orbital (HOMO) on the same molecule. Here the Coulomb
coupling of the electron-hole pair is much larger than the hopping
matrix element of a single hole or electron. Exciton transfer
where the Frenkel exciton concept can be applied, occurs in many
bio-molecules, e.g. rhodopsin, porphyrins, blue copper protein,
carotenoids, and chlorophylls. A well studied molecule is the
light harvesting antenna (LH II) from the bacterial photosystem
Rhodopseudomonas Acidophila. It is characterized by a symmetric
structure and composed of nine identical units forming a ring.
Each unit is composed of a chlorophyll dimer. The light-harvesting
complexes store and transfer excitations with high efficiency.

In the photosynthetic process a LH-II ring absorbs a photon. The
excitation is transferred to other LH-II rings and sent via the
LH-I ring to the reaction center and then converted to chemical
energy. The excitation in the LH-II ring B850 can move over the
whole ring -- it is delocalized over the ring. In other rings,
such as B800, the excitations are usually considered to be more
localized \cite{Fleming}. Furthermore, mechanisms exist which
dissipate excitation energy to safe the organism from damage
\cite{Scholes}. The degree of delocalization depends strongly on
the coupling to the vibronic environment and may be crucial for
the function of the specific protein.

The interpretation of optical spectra \cite{Mukamel} requires a
theory which incorporates both static and dynamic disorder. If the
fluctuations of the protein environment occur on a much larger
time scale than those of the excitonic system, the disorder is
regarded as static. Such a static disorder can be treated by a
thermodynamic average. The dynamic disorder stems from the
coupling of the electronic degrees of freedom to the fluctuations
of the environment. In the present paper we will study the effect
of dissipation while neglecting static disorder.

A full {\it ab initio} quantum chemical calculation of
molecules which show exciton transfer reactions
is impossible; therefore it is reasonable to investigate
the system using simple models which, nevertheless, cover the
relevant physics of the problem. The most elementary non-trivial
model which describes quantum dissipation is the well studied
spin-boson model \cite{Leggett,Weiss}.
It can be viewed as an archetype for modelling
the system-environment interaction in bio-molecules in which the
electronic degrees of freedom couple to a dissipative environment.

A variety of theoretical methods have been developed to calculate
absorption spectra and rates (see e.g.,
\cite{Haken,Barvik,Silbey,Wubs,Agranovich2,Kleinekathoefer1,Kleinekathoefer2,Redfield}).
Some investigations of exciton transfer systems were based on
perturbation theory in the exchange coupling between the excitons
or in the coupling of the electronic system to the vibrations. If
the exciton-vibrational coupling $\alpha$ is weak compared to the
dipole-dipole coupling density-matrix theory is used \cite{May}.
If the intermolecular Coulomb coupling $J$ is small we are in the
limit of nonadiabatic exciton transfer. Here perturbation theory
is applicable which leads to the F{\"o}rster equations \cite{May}.

A key challenge for a theoretical study of exciton transfer is to
cover the whole range of possible behaviour, from coherent to
incoherent transfer or even localization or self-trapping of the
excitations. Here we use the non-perturbative numerical
renormalization group (NRG) to calculate equilibrium properties of
the exciton system in the full parameter space. We give a detailed
study of the phase diagram, dynamic equilibrium properties for
chains and rings up to six sites, and the frequency dependent
linear absorption spectra of excitons in a dimer and trimer
molecule as a function of the coupling to the bosonic bath. The
behaviour of the system is governed by the competition between the
couplings $J$ and $\alpha$ which determine whether the excitations
are delocalized or localized.

The models we are considering here describe general electron-boson
systems with a limited number of quantum states on a few sites
with a coupling to a (quantum) dissipative environment. We
restrict ourselves to donor, acceptor, and bridge molecules with
only two electronic levels per molecule, neglecting the spin
degree of freedom. If the flux of photons is sufficiently low then
the exciton migration in systems such as
a pigment network can be satisfactorily
modelled by a single excitation. We show that in the single-exciton
subspace the multi-site electron-boson model can be mapped to a multi-site exciton-boson
model.
The two-site exciton-boson model is identical to the
spin-boson model. The exciton system is coupled to all degrees of
freedom of the (protein) environment which is modelled by an
infinite set of harmonic oscillators. After the discussion of the
various models in Sec.~II, we introduce the NRG approach used here
in Sec.~III. Section IV is devoted to the results for the phase
diagram, dynamic properties, and the absorption spectrum. We show
how the degree of
delocalization depends on the different Coulomb
interactions, the coupling to the bosonic bath, and the geometric
structure.

\section{Model}
In general we describe the problem by a small electronic system
like a short chain or ring of molecules with the electronic part
$H_{\rm el}$ coupled via the $H_{\rm el-bath}$ part to the
vibronic degrees of freedom incorporated in $H_{\rm bath}$:
\begin{eqnarray}
H=H_{\rm el}+H_{\rm el-bath}+H_{\rm bath} \label{eq1} \ .
\end{eqnarray}
The simplest possible model for the electronic part is a two-state
system with the two states corresponding to the electron being
located at the donor or at the acceptor site. In this case, the
electronic part can be modelled via
\begin{eqnarray}
H_{\rm el}=\sum_{i=A,D} \epsilon_i c_i^{\dag}c_i - t
(c_{A}^{\dag}c_D+ c_{D}^{\dag}c_A) \ .
\end{eqnarray}
Apparently, models of this kind where the electrons are allowed to
hop between donor and acceptor sites (with hopping matrix element
$t$) are connected to electron transfer
problems. In this paper we do not consider such hopping processes
and focus on excitation transfer induced by a two-particle
interaction term in the Hamiltonian. The electronic part of the
Hamiltonian then takes the following general form
\begin{eqnarray}
H_{\rm el}=\sum_{i,l,\sigma} H(i,l,\sigma)+\sum_{i,j;k,l;\sigma,
\sigma'}V(i,j;k,l;\sigma,\sigma') \ , \label{el:model}
\end{eqnarray}
where the first term represents the local part with on-site
energies, on-site Coulomb interactions as well as possible
spin-orbit couplings. The second part describes all non-local
terms, mainly the non-local Coulomb interaction (as a possible
extension also the single-particle hopping term). In principle,
the parameters of the Hamiltonian eq. (\ref{el:model}) can be
extracted from quantum chemical calculations \cite{Rosa,Starikov}.

In the current work, we explore excitons in a chain or ring using
the Hamiltonian
\begin{eqnarray}
H_{\rm el} &=& \sum_{ik } \epsilon_{ik} c_{ik }^{\dag} c_{ik }  +
\sum_{ijlkl'k'} J_{ij} c_{i k}^{\dag} c_{j l}^{\dag} c_{j l' }
c_{i k'} \ .
\label{eq:exciton-general}
\end{eqnarray}
The operators $c_{ik }^{(\dag)}$ denote annihilation (creation)
operators for the electrons on site $i$ in the level $k$,
$\epsilon_{ik}$ is the on-site energy and $J_{ij}$ the exchange
interaction between site $i$ and $j$. The electronic part of each
level is represented by a two-level system. We neglect the
single-electron hopping between neighbouring sites  as well as
spin degrees of freedom. Therefore, the excitations (Frenkel
excitons) can only be transferred via the Coulomb coupling.
Similar models were discussed, e.g.,  in
\cite{Kleinekathoefer2,May,Barvik,Gilmore}

The Hamiltonian eq.~(\ref{eq:exciton-general}) describes a
coherent motion of the excitation through the whole system; this
coherence can be destroyed in the presence of a dissipative
environment. Here, the coupling to the environment is due to the
change of the dipole moment of the molecule during the transition.
Simulations \cite{Schulten2} showed that the coupling involves
essentially all nuclear degrees of freedom of the protein which
have to be described quantum mechanically. Even at physiological
temperatures there are many degrees of freedom in proteins with
frequencies high enough to make a quantum mechanical description
necessary \cite{Schulten2}. We represent the vibrations of the
environment by a set of harmonic oscillators similar to the
spin-boson model.

The last term in eq.~(\ref{eq1}) describes the free bosonic bath
\begin{eqnarray}
H_{\rm bath}=\sum_{n} \omega_{n} b_{n}^{\dagger} b^\pd_{n} \ ,
\end{eqnarray}
with the bosonic annihilation (creation) operators
$b_n^{(\dagger)}$. The second term in eq.~(\ref{eq1}) describes
the coupling of the electrons to the bosonic bath
\begin{eqnarray}
H_{\rm el-bath}= \sum_{i  \sigma} g_{i } n_{i \sigma} \sum_{n}
     \lambda_n     \left(  b_{n}^{\dagger} + b^\pd_{n}  \right) \ ,
\end{eqnarray}
where $\lambda_n$ is the coupling strength to the $n$th
oscillator. We consider here a dipole coupling so that the sum
$\sum_{i} g_{i} n_{i}$ is the polarization operator
of the electronic system.
The values of the parameters
$g_i$ will be specified below.

In analogy to the spin-boson model \cite{Leggett,Weiss}, the
coupling of the electrons to the bath degrees of freedom is
completely specified by the bath spectral function
\begin{equation}
   J(\omega) = \pi \sum_{n}
\lambda_{n}^{2} \delta\left( \omega -\omega_{n} \right) \ .
\label{eq:J}
\end{equation}
Several parametrizations of $J(\omega)$ have been studied in
the literature \cite{Leggett}. For a given system, the bath
spectral function can also be calculated using molecular dynamics
simulations \cite{Kleinekathoefer1}. Here we
restrict ourselves to a simple ohmic spectral function and will use
more realistic spectral functions in a future study.

\subsection{Dimer}

\begin{figure}[t]
\vspace*{0.1cm} \epsfxsize=9cm
\centerline{\epsffile{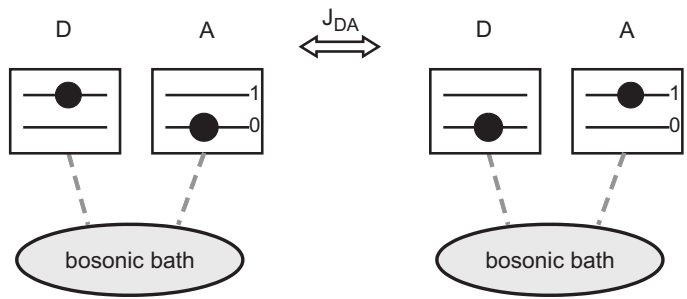}}
 \caption{ Schematic view of
the two-site electron-boson model. The transfer integral for the
exciton transfer between donor (D) and acceptor (A) sites is given
by $J$. Dissipation in the exciton transfer process is due to the
coupling to a common bosonic bath. } \label{model_dimer}
\end{figure}

For the dimer (that is the two-site electron-boson model as sketched
in Fig.~\ref{model_dimer}) the Hamiltonian reduces to
\begin{equation}
   H_{\rm dimer} = H_{\rm el} + H_{\rm el-bath} + H_{\rm bath}  \ ,
\label{eq:dimer}
\end{equation}
with the electronic part
\begin{eqnarray}
H_{\rm el}=\sum_{i=D_0,D_1,A_0,A_1} \varepsilon_{i}
c_{i}^{\dagger}c_{i}+J \left(c_{D_0}^{\dagger} c_{A_1}^{\dagger}
c_{D_1}c_{A_0} + h. c. \right) \label{dimer} \ ,
\end{eqnarray}
and the coupling term
\begin{eqnarray}
H_{\rm el-bath}= \sum_{i=D_0,D_1,A_0,A_1} g_i n_i \sum_{n}
         \lambda_n \left(  b_{n}^{\dagger} + b^\pd_{n}  \right) \ .
\end{eqnarray}

The indices $D_0, A_0$ indicate the ground state on the
donor/acceptor and $D_1, A_1$ the first excited states on the
donor/acceptor. For the $g_{i}$ we put
$g_{D_1}=-g_{D_0}=-g_{A_1}=g_{A_0}=\frac{1}{2}$.

The electronic degrees of freedom in this subspace can be
represented by the four-dimensional basis
\begin{equation}
   \vert i \rangle = \left\{
   \vert 0,0 \rangle,
     \vert 1,0 \rangle,
     \vert 0,1 \rangle,
     \vert 1,1 \rangle
                     \right\}   \ ,
\label{eq:basis-dimer}
\end{equation}
with the notation $\vert {\rm D;A} \rangle$ describing the
donor/acceptor
in the ground state ($D/A=0$) or in the excited state ($D/A=1$).
Introducing the notation
\begin{equation}
    \hat{Y} = \sum_{n} \omega_{n} b_{n}^{\dagger} b^\pd_{n}
    \ \ , \ \
    \hat{X} = \frac{1}{2} \sum_{n}
         \lambda_n \left(  b_{n}^{\dagger} + b^\pd_{n}  \right)
    \ ,
\end{equation}
we arrive at the matrix $M=M_0+M_b= \langle i| H_{\rm dimer} |j
\rangle$ $(i,j=1..4)$, where the matrix elements are taken only
with respect to the electronic states:

\begin{equation}
   M_0=
 \left( \begin{array}{cccc}
         \epsilon_{D_0}+\epsilon_{A_0}  & 0 & 0 & 0\\
         0 & \epsilon_{D_1}+\epsilon_{A_0}& J & 0\\
         0 & J &  \epsilon_{D_0}+\epsilon_{A_1}  & 0 \\
         0 & 0 & 0 & \epsilon_{D_1}+\epsilon_{A_1}
 \end{array}     \right) \ ,
\end{equation}
and
\begin{equation}
   M_b=
 \left( \begin{array}{cccc}
          \hat{Y} & 0 & 0 & 0\\
         0 & \hat{Y}- \hat{X}& 0 & 0\\
         0 & 0 &  \hat{Y}+ \hat{X}  & 0 \\
         0 & 0 & 0 & \hat{Y}
 \end{array}     \right) \ .
\end{equation}

With $\epsilon_{D_0}=\epsilon_{A_0}=0$ and
$\epsilon_{D_1}=\epsilon_{D}$, $\epsilon_{A_1}=\epsilon_{A}$ the
eigenvalues of the electronic part $M_0$ are
\begin{eqnarray*}
E_1 &=& 0 \ ,\\
E_2 &=& \epsilon_D+\epsilon_A \ ,\\
E_{3,4} &=& \frac{\epsilon_{D}+\epsilon_{A}}{2}
\\
&\pm& \sqrt{\frac{ \left(\epsilon_{D}-\epsilon_{A}\right)^2}{4} +
 J^2} \ .
\end{eqnarray*}

The eigenstates with energies $E_{3,4}$ are linear combinations of the
basis states $|1,0\rangle$ and $|0,1\rangle$.

The Hamiltonian of the dimer eq.~(\ref{eq:dimer}) can be
decomposed into subspaces of zero, one, and two excitons. For the
subspace with one exciton in the dimer, the basis
eq.~(\ref{eq:basis-dimer}) reduces to $   \vert i \rangle =
\left\{
     \vert 1,0 \rangle,
     \vert 0,1 \rangle
                     \right\} $
and the Matrix $M$ reads
\begin{equation}
   M=
 \left( \begin{array}{cc}
         \epsilon_A+\hat{Y}- \hat{X} & J   \\
         J & \epsilon_D+\hat{Y}+ \hat{X}
 \end{array}     \right) \ .
 \label{spinboson}
\end{equation}
This matrix allows for an exact mapping onto the spin-boson model
(for a similar discussion, see Ref.~\onlinecite{TTB}). The model
is equivalent to eq. (\ref{multi2}) with $N=2$.

\subsection{Trimer}
\begin{figure}[t]
\vspace*{0.1cm} \epsfxsize=9cm
\centerline{\epsffile{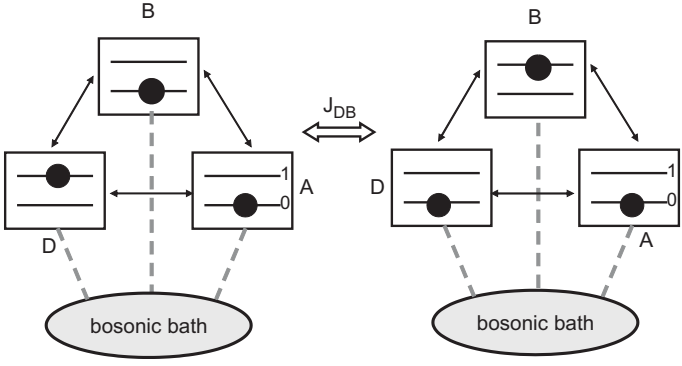}}
 \caption{ Schematic view of
the three-site electron-boson model. The Coulomb matrix element of
excitons between donor (D), bridge (B) and acceptor (A) sites is
given by $J$. Dissipation in the exciton transfer process is due
to the coupling of the electronic degrees of freedom to a
common bosonic bath. The excitation transfer shown in the figure
is due to the coupling $J_{DB}$} \label{model_trimer}
\end{figure}

\begin{figure}[t]
\vspace*{0.1cm} \epsfxsize=9cm
\centerline{\epsffile{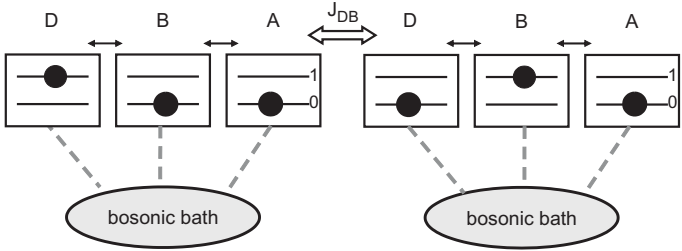}}
 \caption{ Schematic view of
the three-site (chain) electron-boson model. } \label{model_chain}
\end{figure}

The Hamiltonian for a donor-bridge-acceptor system in a
trimer geometry (see Figs.~\ref{model_trimer} and \ref{model_chain})
takes the form
\begin{eqnarray}
H_{\rm el}&=&\sum_{i=D_0,D_1,A_0,A_1,B_0,B_1} \varepsilon_{i}
c_{i}^{\dagger}c_{i} \nonumber \\
&+& J_{DA} \left(c_{D_0}^{\dagger} c_{A_1}^{\dagger}
c_{D_1}c_{A_0} +h.c. \right) \nonumber \\
&+& J_{AB} \left( c_{B_0}^{\dagger} c_{A_1}^{\dagger}
c_{B_1}c_{A_0} +h.c \right) \nonumber
\\ &+& J_{BD} \left(c_{B_0}^{\dagger}
c_{D_1}^{\dagger} c_{B_1}c_{D_0}+ h. c. \right) .
\end{eqnarray}

The electrons are coupled to the bosonic bath via:
\begin{eqnarray}
H_{\rm el-bath}= \sum_{i=D_0,D_1,A_0,A_1,B_0,B_1} g_i n_i
%\left[n_{\rm D_0}-n_{\rm D_1}-n_{\rm A_0}+n_{\rm A_1}
%\right]
\sum_{n}
         \frac{\lambda_n}{2} \left(  b_{n}^{\dagger} + b^\pd_{n}  \right)
\nonumber \\
\end{eqnarray}
For the $g_i$ we choose $g_{D_1}=-g_{D_1}=-g_{A_0}=g_{A_1}=1$ and
$g_{B_0}=g_{B_1}=0$.

The basis for the electronic degrees of freedom is now composed
of eight states:
\begin{eqnarray}
|1\rangle&=&|0,0,0 \rangle, |2\rangle=|1,1,1 \rangle,|3\rangle =
|1,0,0 \rangle, |4\rangle=|0,1,0 \rangle,  \nonumber \\
|5\rangle&=&|0,0,1 \rangle ,|6\rangle=|1,1,0 \rangle,
|7\rangle=|0,1,1 \rangle,|8\rangle= |1,0,1 \rangle \nonumber ,
\end{eqnarray}
where the $0 (1)$ indicates an occupied first (second) level,
respectively, with the notation $|D,B,A \rangle$ for the
occupation of donor, bridge, and acceptor molecule.
A direct hopping of the exciton is possible from the
donor to the acceptor or to the bridge and from the acceptor to
the bridge and back (see Fig.~\ref{model_trimer}).
The matrix-elements now read:

\begin{eqnarray}
\langle 1 |H|1 \rangle&=&\epsilon_{D_0}+\epsilon_{B_0} + \epsilon_{A_0}+\hat{Y}+ \left( g_{B_0}+g_{D_0}+g_{A_0} \right) \hat{X}, \nonumber \\
\langle 2 |H|2\rangle&=&\epsilon_{D_1}+\epsilon_{A_1} + \epsilon_{B_1}+\hat{Y} +\left( g_{A1}+g_{D1}+g_{B1} \right) \hat{X}, \nonumber \\
\langle 3 |H|3 \rangle&=&\epsilon_{A_0}+\epsilon_{B_0}+\epsilon_{D_1} +\hat{Y}+ \left( g_{A_0}+g_{B_0}+g_{D_1} \right) \hat{X}, \nonumber \\
\langle 4 |H|4 \rangle&=&\epsilon_{D_0}+\epsilon_{B_1}+\epsilon_{A_0} +\hat{Y}+ \left( g_{B_1}+g_{D_0} +g_{A_0}\right) \hat{X}, \nonumber \\
\langle 5|H|5 \rangle&=&\epsilon_{D_0}+\epsilon_{A_1}+ \epsilon_{B_0}+\hat{Y}+ \left( g_{D_0}+g_{A_1} +g_{B_0}\right) \hat{X}, \nonumber \\
\langle 6 |H|6 \rangle&=&\epsilon_{A_1}+\epsilon_{B_1} +\epsilon_{D_1}+\hat{Y}+ \left( g_{B_1}+g_{A_1}+ g_{D_1}\right) \hat{X}, \nonumber \\
\langle 7|H|7 \rangle&=&\epsilon_{D_0}+\epsilon_{B_1}+\epsilon_{A_1} +\hat{Y}+ \left( g_{B_1}+g_{D_0}+ g_{A_1}\right) \hat{X}, \nonumber \\
\langle 8 |H|8 \rangle&=&\epsilon_{D_1}+\epsilon_{B_0}+ \epsilon_{A_1} +\hat{Y}+ \left( g_{B_0}+g_{D_1}+g_{A_1} \right) \hat{X,} \nonumber \\
\langle 4|H|5 \rangle&=&\langle 5|H|4 \rangle=J_{AB}, \nonumber \\
\langle 6 |H|8 \rangle&=&\langle 8 |H|6 \rangle=J_{AB}, \nonumber \\
\langle 5 |H|4 \rangle&=&\langle 4 |H|5 \rangle=J_{BD}, \nonumber \\
\langle 7 |H|8 \rangle&=&\langle 8 |H|7 \rangle=J_{BD}, \nonumber \\
\langle 3 |H|5 \rangle&=&\langle 5 |H|3 \rangle=J_{AD}, \nonumber \\
\langle 6 |H|7 \rangle&=&\langle 7 |H|6 \rangle=J_{AD}. \nonumber \\
\end{eqnarray}

For $\epsilon_{A0}=\epsilon_{B0}=\epsilon_{D0}=\epsilon$ and
$J_{AB}=J_{BD}=J_{AD}=J$ the eigenvalues are given by
\begin{eqnarray*}
E_{1} &=& 0, \\
E_2 &=& 3 \epsilon, \\
E_{3} &=& \epsilon -J, \\
E_4 &=& \epsilon +2 J, \\
E_{5} &=& 2 \epsilon -J, \\
E_6 &=& 2 \epsilon +2 J. \\
\end{eqnarray*}
For the chain as in Fig.~\ref{model_chain}, we set
$J_{AD}=0$ and $J_{AB}=J_{BD}=J$. The resulting eigenvalues
are:
\begin{eqnarray*}
E_{1} &=& 0, \\
E_2 &=& 3 \epsilon, \\
E_3 &=& 2 \epsilon, \\
E_4 &=& \epsilon, \\
E_{5} &=& \epsilon \pm \sqrt{2}J, \\
E_6 &=& 2 \epsilon \pm \sqrt{2} J.
\end{eqnarray*}
In the subspace with only one exciton, the basis consists of
the three states $|1, 0, 0 \rangle$, $|0, 1, 0 \rangle$, and
$| 0, 0,1
\rangle$, and the matrix reduces to
\begin{equation}
   M=
 \left( \begin{array}{ccc}
         \epsilon_D+\hat{Y}+ 2 \hat{X} & J_{AB} & J_{AD}  \\
         J_{AB} & \epsilon_B+\hat{Y} & J_{BD} \\
J_{AD} & J_{AB} & \epsilon_A+\hat{Y}-2 \hat{X}
 \end{array}     \right) \ .
\label{trmierboson}
\end{equation}

The eigenvalues for the trimer as in
Fig.~\ref{model_trimer} are
\begin{eqnarray*}
E_{1} &=& \epsilon -J, \\
E_2 &=& \epsilon +2 J, \\
\end{eqnarray*}
and for the chain as in Fig.~\ref{model_chain}
\begin{eqnarray}
E_{1} &=& \epsilon, \nonumber \\
E_2 &=& \epsilon \pm \sqrt{2} J. \nonumber
\end{eqnarray}
The model is equivalent to eq. (\ref{multi2}) with $N=3$.
\subsection{Multi-site exciton-boson model}

In the single-exciton subspace, the fermionic degrees of freedom
of the models introduced above can be mapped onto operators $a_i^{(\dagger)}$
for hard-core bosons corresponding to the creation and annihilation
of an exciton at site $i$. This results in general
multi-site exciton-boson models with $N$ sites defined by
\begin{equation}
H_{\rm multi}=H_{\rm x}+H_{\rm x-bath}+H_{\rm bath} ,
\end{equation}
with the electronic part defined as
\begin{eqnarray}
H_{\rm x}= \sum_{i,j}^N J_{ij} a_{i}^{\dagger} a_{j} \label{multi} \ .
\end{eqnarray}
The parameters $J_{ij}$ for $i \neq j$ are the transfer integrals between
site $i$ and $j$. As before, we only consider nearest neighbour
interactions. The diagonal elements $J_{ii}$ are the on-site
energies $\varepsilon_i$ at site $i$. We perform a constant shift
of the Hamiltonian by $J_{ii}=J_{jj} (\forall i,j)$ and arrive at
\begin{eqnarray}
H_{\rm x}=  \sum_{i,j, i \neq j}^N J_{ij}  a_{i}^{\dagger} a_{j}
\label{multi2} \ .
\end{eqnarray}

For the coupling term we assume the following form
\begin{eqnarray}
H_{\rm x-bath}= \sum_{i}^N g_i a_i^{\dagger} a_i \sum_{n} \lambda_n
\left( b_{n}^{\dagger} + b^\pd_{n}  \right) \label{multi_boson}
\end{eqnarray}
with $g_i=(i-(N+1)/2)$. For $N=2$ the
model is equivalent to the spin-boson model with the matrix $M$ as
in eq.~(\ref{spinboson})
and for $N=3$ equivalent to eq.~(\ref{trmierboson}).

The eigenvalues of $H_{\rm x}$ for $N=4$  are
$E=0,\pm 2 J$ for the ring geometry and $E=\pm J$ for the chain.
The eigenvalues for the ring with $N=5$ are
$E=2J,(\sqrt{5}-1)J/2,(-\sqrt{5}-1)J/2$ and for the
ring with $N=6$: $E=\pm J, \pm 2 J$.

\section{Method}

The models we are considering here are completely specified by the
parameters of the electronic system and the spectral function
$J(\omega)$ (defined in eq.~(\ref{eq:J})) which can be estimated
in a classical molecular dynamics simulation. We are using here an
ohmic form:
\begin{equation}
   J(\omega) = 2 \pi \alpha \omega \Theta (\omega - \omega_c) \ ,
\end{equation}
where $\alpha$ is the dimensionless coupling
for which we use values in the range 0.01-2.
The parameters J
and t are all measured in units of $\omega_c$.
Typical values of $\hbar \omega_c$ are
of the order of 1 to 10 meV.

As described in the introduction, basically all degrees of freedom of the
bosonic bath (the dissipative environment) are relevant for
the behaviour of the electronic or excitonic system. So it is
not possible to disregard high energy states even if we are
interested in  low temperature properties like the coherent
behaviour for temperatures smaller than the characteristic
temperature $T^*$.
The renormalization group ansatz is designed for problems where
every energy scale contributes and
perturbation theory typically shows logarithmic divergencies at small
frequencies (energies) when the temperature goes to zero.

In order to keep the paper self– contained we explain the
Numerical Renormalization Group (NRG) method for the bosonic bath
in detail.

Originally the NRG was invented by Wilson for a fermionic bath to
solve the Kondo problem \cite{Wil75,Kri80}. The fermionic NRG is a
standard and powerful tool to investigate complex impurity
problems with one or more fermionic baths. Only recently, the
method was extended to treat quantum impurity systems with a
coupling to a {\em bosonic} bath \cite{BLTV,BTV}. Here we focus on
equilibrium quantities (recently it was shown that the NRG can
also be applied to non-equilibrium situations
\cite{AndersSchiller}.)

\begin{figure}[t]
\vspace*{0.1cm} \epsfxsize=9cm
\centerline{\epsffile{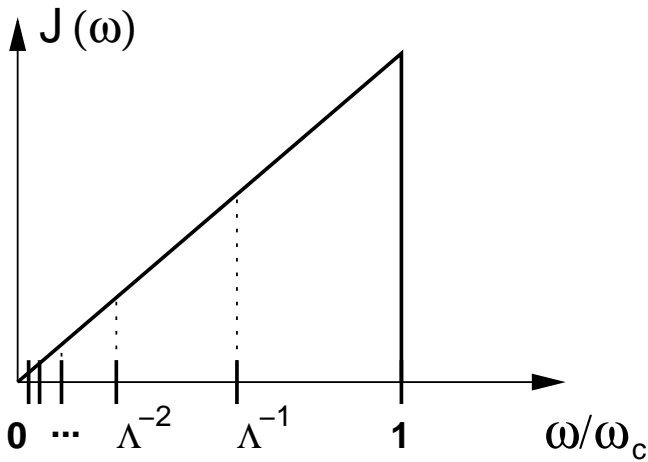}}
\caption{Logarithmic discretization of the spectral function
$J(\omega)$.} \label{fig:dis}
\end{figure}

\begin{figure}[t]
\vspace*{0.1cm} \epsfxsize=5cm
\centerline{\epsffile{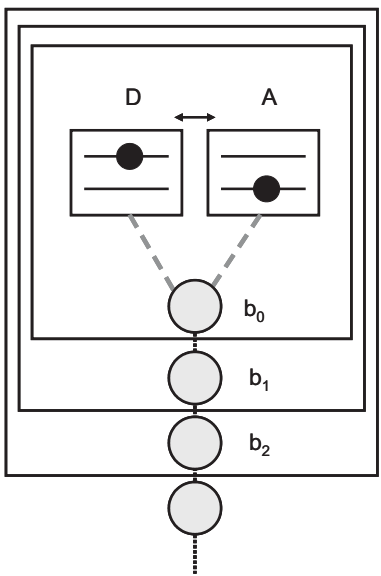}}
\caption{Scheme of the bosonic chain-NRG. The boxes represent the
iterative diagonalization.} \label{fig:abs:dimer3}
\end{figure}

For the numerical renormalization procedure we start from the
Hamiltonian written in a continous form
(see the detailed discussion in Ref.~\onlinecite{BLTV}):

\begin{eqnarray}
H =H_{\rm el}+\int_0^1 d\varepsilon g(\varepsilon)
b_{\varepsilon}^{\dag} b_{\varepsilon} + \left(\sum_i g_i
n_i\right) \int_0^1 d\varepsilon (b_{\varepsilon}^{\dag}+
b_{\varepsilon})\ . \nonumber \\
\end{eqnarray}
The function $g(\varepsilon)$ is the dispersion of the phonon bath
and $h(\varepsilon)$ characterizes the coupling between the
electronic system and the  bath. Both functions are related to the
spectral function $J(\omega)$, see Ref.~\onlinecite{BLTV}. We
start defining the renormalization group transformations by a
logarithmic discretization of the bath spectral function
(Fig.~\ref{fig:dis}) in intervals $[\Lambda^{-n+1},\Lambda^{-n}]$,
with $n=0,1,\ldots,\infty$ and $\Lambda>1$ the NRG discretization
parameter. The discretization is exact for $\Lambda \rightarrow 1$
and works still very well for $\Lambda=3$. (Here we are using
$\Lambda=2$)

Within each of these intervals only one bosonic degree of freedom
is retained as a representative of the continuous set of degrees
of freedom. The function $h(\varepsilon)$ is chosen to be a
constant in each intervall of the logarithmic discretization. The
Hamiltonian is written in the new discrete basis and the
resulting Hamiltonian is mapped onto a semi-infinite chain
(Fig.~\ref{fig:abs:dimer3}) with the electronic part
$H_{\rm el}$ coupling to the
first site of the bosonic chain. Finally the chain-Hamiltonian is
numerically diagonalized via successively adding one site to the
chain. The effective Hamiltonian is treated on successive smaller
energy scales by the renormalization group transformation
\begin{eqnarray}
H_{N+1}&=&\Lambda H_N \nonumber \\
&+&\Lambda^{N+1} \left[ \varepsilon_{N+1} b_{N+1}^{\dag} b_{N+1}
+t_{N} \left( b_{N}^{\dag} b_{N+1} +h.c. \right)\right]. \nonumber \\
\end{eqnarray}
The energies $\varepsilon_n$ and couplings between the elements of
the chain are falling off as $\Lambda^{-n}$.

The bosonic NRG has been shown to give very accurate results for
the spin-boson model \cite{BTV,BLTV}. One of its strengths is the
flexibility to handle a variety of models involving the coupling
of a small subsystem to a bosonic bath.

To detect possible phase transitions, we calculate the eigenvalue
spectrum and the density-density correlation function (see below).
In the limit of $\alpha=0$ the exciton system and the bosonic
degrees of freedom are completely decoupled.
The coherent motion of the exciton is
undamped and we are in the delocalized phase. In
contrast, in the case of $J=0$ the
system is in the localized phase. The two phases (localized
and delocalized) are connected by a quantum phase
transition. Similar to the analysis in Ref.~\onlinecite{BLTV},
the phase diagrams of the exciton-boson models studied here
can be obtained from the flow diagram of the lowest-lying
many-particle levels. Another possibility is to calculate the density-density
autocorrelation function $C(\omega)$ which shows a divergency at
the phase transition.

 We calculate $C(\omega)$ for the dimer and trimer for different
sets of parameters. This quantity is defined by
\begin{eqnarray}
C(\omega) = \frac{1}{4 \pi} \int_{-\infty}^{\infty} dt e^{i \omega
t} \left\langle
 \left[ \sum_i g_i a^{\dag}_i a_i(t), \sum_i g_i a^{\dag}_i a_i  (0) \right] \right\rangle \ ,
 \nonumber
\end{eqnarray}
and probes the dynamics under equilibrium preparation.
For the two-site model, $C(\omega)$ corresponds to the spin-spin correlation
function of the equivalent spin-boson model
\begin{eqnarray}
C(\omega) = \frac{1}{4 \pi} \int_{-\infty}^{\infty} dt e^{i \omega
t} \left\langle
 \left[ \sigma_z(t), \sigma_z  (0) \right] \right\rangle
 \nonumber ,
\end{eqnarray}
with $\sigma_z$ the $z$-component of the spin in this model.

We calculate the density-density correlation function as sum of
$\delta$-functions in the Lehmann representation:
\begin{eqnarray}
& & C(\omega) = \frac{1}{2} \sum_n \left|\left\langle 0\left|\sum_i g_i
a^{\dag}_i a_i \right|n \right \rangle \right|^2 \delta \left(
\omega+\epsilon_0-\epsilon_n \right),\nonumber  \\
\ \ & & \omega > 0 \ .
\end{eqnarray}
The linear absorption and emission coefficient $\alpha(\omega)$
for the donor site of the electronic system coupled to the bosonic
bath under influence of an external laser field of frequency
$\omega$ is given by Fermi's golden rule:

\begin{eqnarray}
& & \alpha^{D}(\omega)=2\pi \sum_{f} \left| \left\langle
  f \left|H_{\rm pert}^{D}\right|0\right\rangle\right|^2 \delta\left(\omega +E_0-E_f
  \right),  \nonumber \\ \label{eq:alphaD}
\end{eqnarray}
where $H^D_{\rm pert}$ is defined as
\begin{eqnarray}
 H_{pert}^{D}&=&c_{D1}^{\dag} c_{D0} + c_{D0}^{\dag} c_{D1}.
\end{eqnarray}
The term $H_{\rm pert}^{D}$ describes the excitation of an electron from the
ground state $D_0$ to the excited state $D_1$. It can be treated
perturbatively as long as the probing photon energy is small. For
the initial state  we use the ground state $|0\rangle$
and $|f\rangle$ are all possible final states.

The eigenenergies of $H$ ($E_0$ and $E_f$) and the matrix elements
$\langle 0|H_{\rm pert}|f \rangle$ are evaluated with the NRG
 for different $J$ and increasing coupling to
the bosonic bath.
To obtain a continuous curve for $\alpha^D(\omega)$, the
$\delta$-functions appearing in eq.~\ref{eq:alphaD} have
to be broadened. Here we use the strategy discussed in
Ref.~\onlinecite{BCV}, that is replacing each $\delta$-function
by a gaussian on a logarithmic scale. On a linear scale,
this function is not symmetric around its center so that
spectral weight in $\alpha^D(\omega)$ appears to be
shifted to higher frequencies.

\section{ Results}

\subsection{Phase Diagram}

Increasing $J$ tends to delocalize the exciton. Since we
have excluded the
single-electron hopping in the Hamiltonian,
the dynamics can be restricted to
the single-exciton subspace which maps onto a spin-boson model
for two sites.  For increasing coupling to the bosonic bath, the exciton
localizes at a critical $\alpha_{\rm c}$.
We explore the phase diagram with localized and
delocalized phases (connected by a Kosterlitz-Thouless transition)
for the exciton-boson model with 2 to 6 molecular sites by
calculating the critical $\alpha_c (J)$ as a function of $J$.
Note that for a finite bias $\varepsilon$ (difference of the
excitation energy on donor and acceptor) there is no phase
transition but a crossover from a delocalized to a more localized
regime.

The phase diagram for the spin-boson model was calculated already
in Ref.~\onlinecite{BLTV}. The critical $\alpha$ depends linearly on the
matrix element $J$. It was noted that the exact value of
$\alpha_c$ has to be determined in the limit of $\Lambda
\rightarrow 1$. We do not perform the extrapolation so that the
critical $\alpha_c$ is somewhat larger than the actual value.

\begin{figure}[t]
\vspace*{0.1cm} \epsfxsize=9cm
\centerline{\epsffile{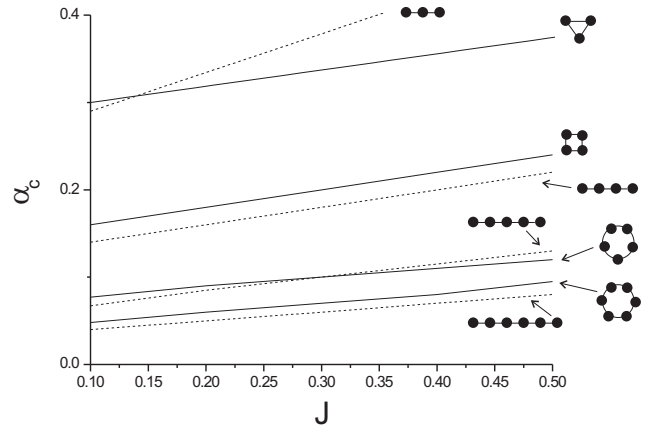}}
\caption{Phase diagram of rings (solid line) and chains (dotted
line) with 3,4,5, and 6 sites. The system is in the localized
(delocalized) phase above (below) the phase boundary. }
\label{fig:phase:large}
\end{figure}

In Fig.~\ref{fig:phase:large}
we display the phase diagram of the
multi-site exciton-boson model for 3,4,5 and 6 sites for both
chain and ring geometry.

 The dashed and solid lines
display the critical $\alpha_c$ for the chain and ring,
respectively. The critical coupling shows a linear behaviour
similar as in the spin-boson model. For an even number of sites the
ring has a larger critical $\alpha$ than the chain. For an odd
number of sites, both curves cross at a certain value of $J$
above which the
opening of the ring will tends to delocalize the exciton.

For a multi-site exciton-boson model with three or more sites, no
quantum phase transition is observed as soon as the couplings
$J_{ij}$ between neighbouring sites are different. To study the
crossover from the delocalized to a more localized phase, we
calculate equilibrium dynamical properties as discussed in the
following.

\subsection{Equilibrium Dynamical Properties}

\begin{figure}[t]
\vspace*{0cm} \epsfxsize=9cm
\centerline{\epsffile{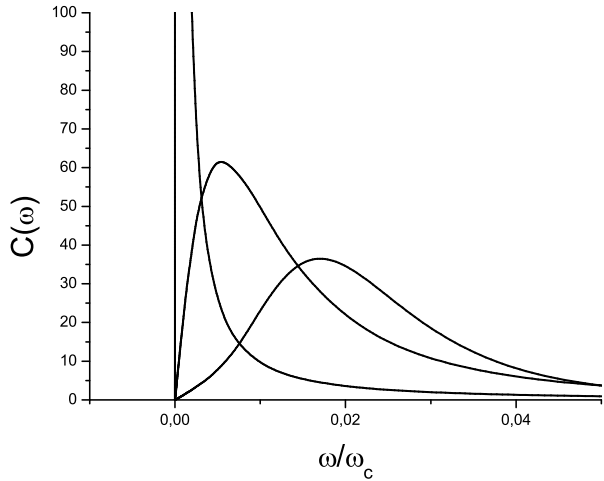}}
 \caption{Density-density correlation function $C(\omega)$ for
the dimer for different $\alpha$ and $J=0.4$. For small
frequencies, $C(\omega)$ is
linear in $\omega$. The slope increases with increasing $\alpha=0.1,0.2,0.3$.
For zero bias ($\varepsilon = 0$) the phase transition to the
localized phase is indicated by an infinite slope.}
\label{fig:phase:trimer}
\end{figure}

To study the dynamics of the electron transfer process in the
one-exciton subspace we calculate the density-density correlation
function $C(\omega)=\frac{1}{2\pi} \int_{-\infty}^{+\infty} e^{i
\omega t} C(t)\, {\rm d}t$ with
\begin{equation}
C(t) \sim \left \langle [a_D^{\dag} a_D(t)-a_A^{\dag}
a_A(t),a_D^{\dag} a_D(0)-a_A^{\dag} a_A(0)]_+ \right \rangle,
\end{equation}
for the dimer ($g_D=1/2, g_A=-1/2$) and for the trimer
($g_B=0,g_A=-1,g_D=1$).

In the two-site case, the density-density correlation function is
identical to the spin-spin correlation function. The correlation
function shows the power-law behaviour for low frequencies up to
$\omega\approx T^{*}$. When $\alpha$ approaches $\alpha_{\rm c}$,
the slope in $C(\omega)$ increases and the peak position (the
characteristic energy and temperature scale $T^\ast$) is shifted
to lower energies, see Fig.~\ref{fig:phase:trimer}. At the phase
transition the correlation function is diverging. The correlation
function shows an algebraic long time behaviour for $T=0$ and an
exponential decay for finite $T$.

\begin{figure}[t]
\vspace*{0.1cm} \epsfxsize=9cm
\centerline{\epsffile{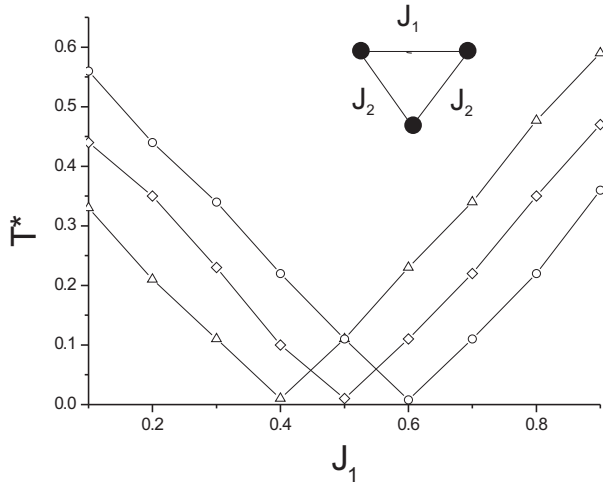}}
 \caption{Crossover
temperature for a trimer ring with $J_2=J_{AB}=J_{DB}=0.4$
(tiangles), $0.5$ (diamonds), $0.6$ (circles) and various values
of $J_1$. The coupling $\alpha$ is set to $0.1$}\label{Ts1}
\end{figure}

\begin{figure}[t]
\vspace*{0.1cm} \epsfxsize=9cm
\centerline{\epsffile{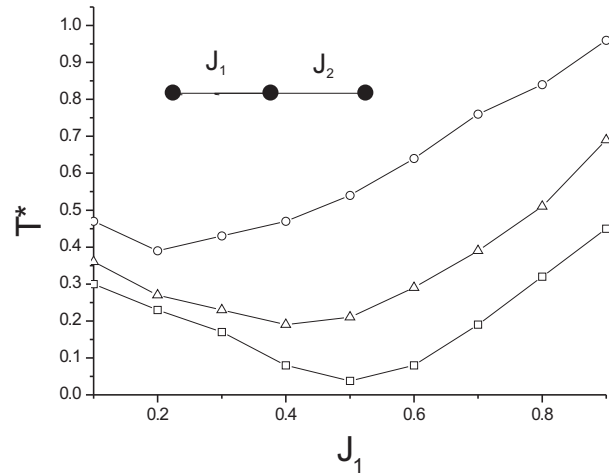}}
 \caption{Crossover
temperature for a trimer chain with $J_{2}=0.5$ and various values
of $J_{1}$. The coupling $\alpha$ is set to $0.1$,$0.2$ and $0.3$
for the upper, middle and lower curve, respectively} \label{Ts2}
\end{figure}

In Fig.~\ref{Ts1} we show results for the characeristic temperature for
a trimer ring  by keeping $J_{AB}$ and $J_{DB}$ constant and
varying $J_1=J_{DA}$. ($J_{DA}=0$ corresponds to the chain.).
The value of $T^*$ goes through a minimum at
$J_1=J_{AB}=J_{DB}$. The larger the difference
the larger is the characteristic temperature. In Fig.~\ref{Ts2} we
display the characteristic temperature for an asymmetric three-site
chain
with $J_1 \neq J_2$ for increasing $\alpha$. For large $\alpha$
and $J_1 = J_2$ the characteristic temperature $T^*$ goes to zero
indicating the phase transition. For the asymmetric chain no phase
transition occurs and $T^*$ increases with the increasing
difference of the matrix elements $J_1$ and $J_2$.

\subsection{Absorption Spectrum}

For $\alpha=0$ the spectrum of the two-site electron-boson
model consists of four states. In the ground state, both
electrons occupy the lowest level of donor and acceptor molecules,
respectively. The system can be excited by a photon:
$D+A \rightsquigarrow D^*+A $. If we now consider a finite $J$,
the exciton is able to move to the acceptor and back ($
D^*+A \leftrightarrows D+A^*$).

\begin{figure}[t]
\vspace*{-0.5cm} \epsfxsize=9cm
\centerline{\epsffile{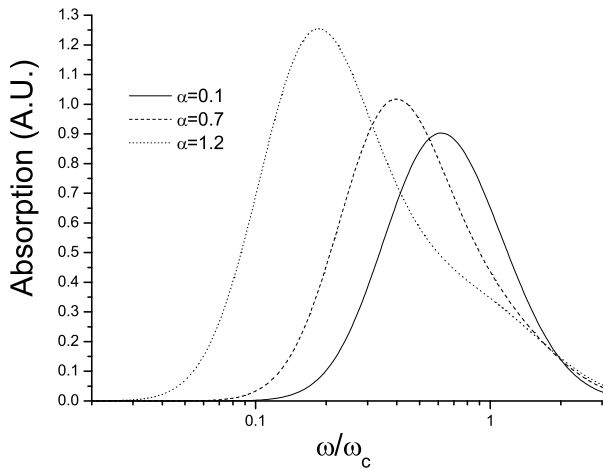}}
 \caption{Dimer absorption
spectrum for $J=0.2$, $\epsilon_D=\epsilon_A=0.75 \omega_c$ and
$\alpha=0.1,0.7,1.2$ as a function of $\omega$.} \label{abs1_don}
\end{figure}

\begin{figure}[t]
\vspace*{0.1cm} \epsfxsize=9cm
\centerline{\epsffile{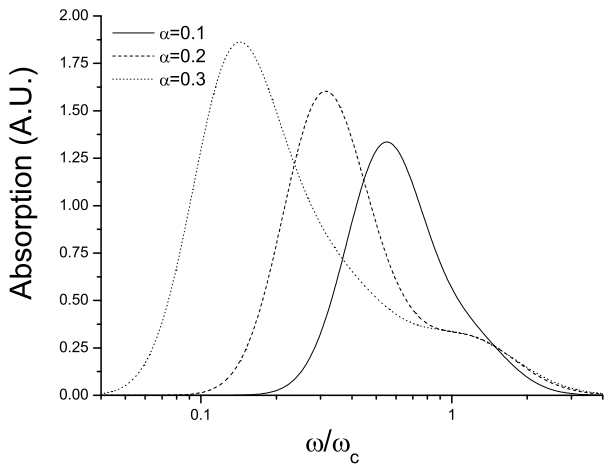}}
 \caption{Absorption
spectrum of a trimer ring for $J=0.2$, $\epsilon_D=\epsilon_A=0.75
\omega_c$ and $\alpha=0.1,0.2,0.3$ as a function of $\omega$. }
\label{abs2_don}
\end{figure}

\begin{figure}[t]
\vspace*{0.1cm} \epsfxsize=9cm
\centerline{\epsffile{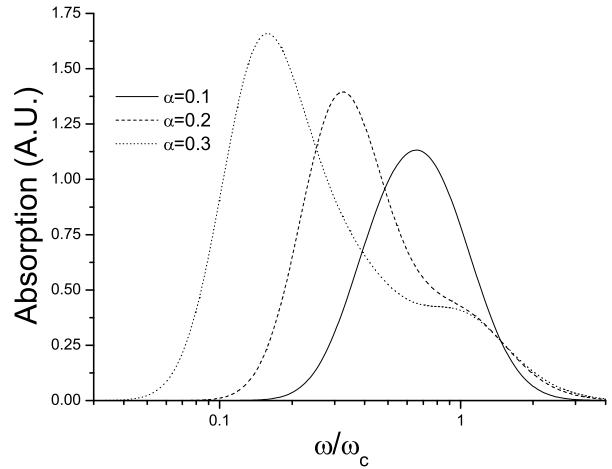}}
 \caption{ Chain absorption
spectrum for $J=0.3$, $\epsilon_D=\epsilon_A=0.75 \omega_c$ and
$\alpha=0.1,0.2,0.3$.} \label{dimer_eps}
\end{figure}

To calculate the absorption spectrum we choose the intitial state
to be the ground state. The ground state is calculated with the
NRG and depends on $\alpha$ and $J$. We set the energy difference
between the ground state and the excited state to
$\varepsilon=0.75 \omega_c$. For $\alpha=0$ and $J=0$ the peak in
the absorption spectrum is at $\omega=\varepsilon=0.75$. For
increasing $J$ peaks are at frequencies equal to the eigenenergies
$\omega=\varepsilon \pm J$. If now $\alpha$ is increased the two
main peaks are broadened and shifted (see Fig.~\ref{abs1_don}).
The height of the peak at low frequencies increases with
increasing $\alpha$.

The absorption spectrum for the trimer with $\alpha=0$
shows peaks at $\varepsilon+2J$ and $\varepsilon-J$
for the ring geometry (with $J_{AB}=J_{AD}=J_{BD}=J$)
and at $\varepsilon \pm \sqrt{2}J$ for the chain.
The absorption spectra for various values of $\alpha$ are shown in
Figs.~\ref{abs2_don} and \ref{dimer_eps} for the ring and chain,
respectively.

\section{Conclusion}

In this paper we studied the phase diagram, equilibrium
dynamical properties
and the linear absorption spectrum of Frenkel excitons
in various ring and chain models with a coupling of the
electronic degrees of freedom to a bosonic bath. We used
the numerical renormalization group method which allows to study
the electron-boson and exciton-boson models in the full parameter regime.

We studied in detail the phase diagrams of the multi-site
electron-boson models in the subspace of one exciton. In the
zero-bias case (all molecules have degenerate HOMO and LUMO energies), increasing the value of $\alpha$ leads to
a quantum phase transition between a delocalized
and a localized phase. For the two-site case (dimer)
the exciton-boson model can be mapped onto the spin-boson
model for which the phase diagram is already known.
For more than two sites, the behaviour is more complicated
and depends also on the geometry (chain vs.~ring).

The calculation of the density-density correlation function
allows to estimate the characteristic temperature for the
crossover between delocalized and localized phase. This crossover
temperature $T^*$ is zero for the
localized phase and increases when the system goes to the
delocalized phase. Increasing the difference between the various
couplings of the model generally leads to a more delocalized behaviour.

It would be interesting to compare our results to optical
experiments of small bio-engineered systems or quantum dots
in which exciton transfer occurs. Further studies are
planned to evaluate the time dependent behaviour of excitons, to
extend the system to larger rings and to include static disorder
using time dependent and equilibrium NRG to model systems like the
LH II ring.

\section*{Acknowledgement}
We acknowledge helpful discussions with U. Kleinekath{\"o}fer,
M. Wubs, and D. Vollhardt. This research was supported by the
DFG through SFB 484 (ST, RB), the Center for Functional
Nanostructures (NT), and by the Alexander von Humboldt foundation
(NT).

%%%%%%%%%%%%%%%%%%%%%%%%%%%%%%%%%%%%%%%%%%%%%%%%%%%%%%%%%%%%%%%%%%%%%%%%%

\end{document}